\documentclass[12pt]{article}
\usepackage{amsfonts}
\usepackage{myart}

\input tcilatex
\begin{document}

\title{Generalization of the relativity theory on the arbitrary space-time
geometry}
\author{Yuri A.Rylov}
\date{Institute for Problems in Mechanics, Russian Academy of Sciences,\\
101-1, Vernadskii Ave., Moscow, 119526, Russia.\\
e-mail: rylov@ipmnet.ru\\
Web site: {$http://rsfq1.physics.sunysb.edu/\symbol{126}rylov/yrylov.htm$}\\
or mirror Web site: {$http://gasdyn-ipm.ipmnet.ru/\symbol{126}%
rylov/yrylov.htm$}}
\maketitle

\begin{abstract}
Contemporary relativity theory is restricted in two points: (1) a use of the
Riemannian space-time geometry and (2) a use of inadequate (nonrelativistic)
concepts. Reasons of these restrictions are analysed in \cite{R2010}.
Eliminating these restrictions the relativity theory is generalized on the
case of non-Riemannian (nonaxiomatizable) space-time geometry. Taking into
account a progress of a geometry and introducing adequate relativistic
concepts, the elementary particle dynamics is generalized on the case of
arbitrary space-time geometry. A use of adequate relativistic concepts
admits one to formulate the simple demonstrable dynamics of particles.
\end{abstract}

\section{Introduction}

Necessity of the general relativity generalization arises as a result of a
geometry progress \cite{R2001}. Now we know nonaxiomatizable (physical)
geometries, which were unknown 20 years ago. Physical geometries are
essentially the metric geometries, whose metric is free of almost all
conventional restrictions. In a metric geometry there exists a problem, how
one should define geometric concepts and rules of geometric objects
construction. One can construct sphere and ellipsoid, which are defined in
terms of metric (world function). However, one needs to impose constraints
on metric (triangle axiom) even for construction of a straight line. It is
unclear, how one can define the scalar product and linear dependence of
vectors. The deformation principle \cite{R2007} solves the problem of the
geometrical concepts definition, without imposing any restrictions on the
metric. The physical geometry is equipped by the deformation principle,
which admits one to construct all definitions of the physical geometry as a
deformation of corresponding definitions of the proper Euclidean geometry.

In the physical geometry the information on the geometry dimension and its
topology appears to be redundant. It is determined by the metric (world
function \cite{S60}), and one may not give it independently. A physical
geometry is described completely by its world function. The geometry is
multivariant and nonaxiomatizable in general. The world function describes
uniformly continuous and discrete geometries. As a result the dynamic
equations in physical space-time geometry are finite difference (but not
differential) equations. Besides, in the physical space-time geometry the
particle dynamics can be described in the coordinateless form. It is
conditioned by a possibility of ignoring the linear vector space, whose
properties are not used in a physical geometry. It is rather uncustomary for
investigators dealing with the Riemannian geometry, which is based on usage
of the linear vector space properties.

There is only one uniform isotropic geometry (geometry of Minkowski) in the
set of Riemannian geometries, whereas there is a lot of uniform isotropic
geometries among physical geometries. In particular, let us consider the
world function $\sigma $ of the form%
\begin{equation}
\sigma =\sigma _{\mathrm{M}}+\lambda _{0}\mathrm{sgn}(\sigma _{\mathrm{M}%
}),\qquad \lambda _{0}=\frac{\hbar }{2bc}  \label{a1.1}
\end{equation}%
where $\sigma _{\mathrm{M}}$ is the world function of Minkowski, and $\hbar
,c,b$ are respectively quantum constant, the speed of the light and some
universal constant. The space-time geometry is discrete and multivariant.
Free particle motion appears stochastic (multivariant). Its statistical
description is equivalent to quantum description in terms of the Schr\"{o}%
dinger equation \cite{R91}.

Thus, application of the physical geometry in the microcosm admits one to
give a statistical foundation of quantum mechanics and to convert the
quantum principles into appearance of the correctly chosen space-time
geometry. One should expect, that a consideration of a more general
space-time geometry and a refusal from the Riemanniance, which is
conditioned by our insufficient knowledge of geometry, will lead to a
progress in our understanding of gravitation and cosmology.

An arbitrary space-time geometry is described completely by the world
function $\sigma (P,P^{\prime })$, given for all pairs of points $%
P,P^{\prime }$. Information on dimension and on topology of the geometry is
redundant, as far as it may be obtained from the world function. The
Riemannian geometry, which is used in the contemporary theory of
gravitation, is considered usually to be the most general possible
space-time geometry. However, it cannot describe a discrete geometry, or a
geometry, having a restricted divisibility. The world function of the
Riemannian geometry satisfies the equation%
\begin{equation}
\frac{\partial \sigma }{\partial x^{i}}g^{ik}\left( x\right) \frac{\partial
\sigma }{\partial x^{k}}=2\sigma ,\qquad \sigma \left( x,x^{\prime }\right)
=\sigma \left( x^{\prime },x\right)  \label{a1.2}
\end{equation}%
It means, that in the expansion%
\[
\sigma \left( x,x^{\prime }\right) =\frac{1}{2}g_{ik}\left( x\right) \xi
^{i}\xi ^{k}+\frac{1}{6}g_{ikl}\left( x\right) \xi ^{i}\xi ^{k}\xi
^{l}+...\qquad \xi ^{k}=x^{k}-x^{\prime k} 
\]%
the metric tensor determines completely the whole world function.

Conventional gravitation equations determine only metric tensor. The world
function and the space-time geometry are determined on the basis of
supposition on the Riemannian geometry. Generalization of the gravitation
equations admits one to obtain the world function directly (but not only the
metric tensor).

The deformation principle admits one to construct all definitions of a
physical geometry as a result of deformation of definitions of the proper
Euclidean geometry. One uses the fact, that the proper Euclidean geometry $%
\mathcal{G}_{\mathrm{E}}$ is an axiomatizable geometry and a physical
geometry simultaneously. It means, that all definitions of the Euclidean
geometry, obtained in the framework of Euclidean axiomatics can be presented
in terms and only in terms of the world function $\mathcal{\sigma }_{\mathrm{%
E}}$ of the Euclidean geometry $\mathcal{G}_{\mathrm{E}}$. Replacing $%
\mathcal{\sigma }_{\mathrm{E}}$ in all definitions of the Euclidean geometry 
$\mathcal{G}_{\mathrm{E}}$ by a world function $\sigma $ of some other
geometry $\mathcal{G}$, one obtains all definitions of the geometry $%
\mathcal{G}$. Definition of the scalar product $\left( \mathbf{P}_{0}\mathbf{%
P}_{1}.\mathbf{Q}_{0}\mathbf{Q}_{1}\right) $ of two vectors $\mathbf{P}_{0}%
\mathbf{P}_{1}$ and $\mathbf{Q}_{0}\mathbf{Q}_{1}$ and their equivalency $%
\left( \mathbf{P}_{0}\mathbf{P}_{1}\text{eqv}\mathbf{Q}_{0}\mathbf{Q}%
_{1}\right) $ are the most used definitions%
\begin{equation}
\left( \mathbf{P}_{0}\mathbf{P}_{1}.\mathbf{Q}_{0}\mathbf{Q}_{1}\right)
=\sigma \left( P_{0},Q_{1}\right) +\sigma \left( P_{1},Q_{0}\right) -\sigma
\left( P_{0},Q_{0}\right) -\sigma \left( P_{1},Q_{1}\right)  \label{a1.3}
\end{equation}%
\begin{equation}
\left( \mathbf{P}_{0}\mathbf{P}_{1}\text{eqv}\mathbf{Q}_{0}\mathbf{Q}%
_{1}\right) \ \text{ if }\left( \mathbf{P}_{0}\mathbf{P}_{1}.\mathbf{Q}_{0}%
\mathbf{Q}_{1}\right) =\left\vert \mathbf{P}_{0}\mathbf{P}_{1}\right\vert
\cdot \left\vert \mathbf{Q}_{0}\mathbf{Q}_{1}\right\vert \wedge \left\vert 
\mathbf{P}_{0}\mathbf{P}_{1}\right\vert =\left\vert \mathbf{Q}_{0}\mathbf{Q}%
_{1}\right\vert  \label{a1.4}
\end{equation}

They are defined in such a way in the Euclidean geometry. They are defined
in the same way also in any physical geometry.

Solution of (\ref{a1.4}) is unique in the case of the proper Euclidean
geometry, although there are only two equations, whereas the number of
variables to be determined is larger, than two. For arbitrary physical
geometry a solution is not unique, in general. As a result there are many
vectors at the point $P_{0}$, which are equivalent to vector $\mathbf{Q}_{0}%
\mathbf{Q}_{1}$ at the point $Q_{0}$. Even geometry of Minkowski is
multivariant with respect to spacelike vectors, although it is
single-variant with respect to timelike vectors. Space-time geometry becomes
to be multivariant with respect to timelike vectors only after proper
deformation.

\section{Influence of the matter distribution on \newline
space-time geometry}

At the generalization of the general relativity on the case of arbitrary
space-time geometry the two circumstances are important.

\begin{enumerate}
\item A use of the deformation principle,

\item A use of adequate relativistic concepts, in particular, a use of
relativistic concept of the events nearness (See details in \cite{R2009}).
\end{enumerate}

Two events $A$ and $B$ are near, if and only if 
\begin{equation}
\sigma \left( A,B\right) =0  \label{a1.5}
\end{equation}%
In the space-time of Minkowski a variation $\delta g_{ik}$ of the metric
tensor under influence of the matter have the form%
\begin{equation}
\delta g_{ik}\left( x\right) =-\kappa \int G_{\mathrm{ret}}\left(
x,x^{\prime }\right) T_{ik}\left( x^{\prime }\right) \sqrt{-g\left(
x^{\prime }\right) }d^{4}x^{\prime },\quad \kappa =\frac{8\pi G}{c^{2}}
\label{a1.6}
\end{equation}%
where $T_{ik}$ is the energy-momentum tensor of the matter,%
\begin{equation}
G_{\mathrm{ret}}\left( x,x^{\prime }\right) =\frac{\theta \left(
x^{0}-x^{\prime 0}\right) }{2\pi c}\delta \left( 2\sigma _{\mathrm{M}}\left(
x,x^{\prime }\right) \right) ,\qquad \theta \left( x\right) =\left\{ 
\begin{array}{ccc}
1 & \text{if} & x>0 \\ 
0 & \text{if} & x<0%
\end{array}%
\right.  \label{a7}
\end{equation}%
and $G$ is the gravitational constant.

Appearance of world function in the $\delta $-function means, that the
condition of the nearness $\sigma _{\mathrm{M}}(x,x^{\prime })=0$ leads to
interpretation of gravitational (and electromagnetic) interactions as a
direct collision of particles. Being presented in terms of world function
these formulae have the same form in any physical geometry.%
\begin{eqnarray}
\delta \sigma \left( S_{1},S_{2}\right) &=&-G\dsum\limits_{s}m_{s}\frac{%
\theta \left( \left( \mathbf{P}_{l}^{\prime }\mathbf{P.PQ}_{0}\right)
\right) }{\left( \mathbf{P}_{l}^{\prime }\mathbf{P.P}_{l}^{\prime }\mathbf{P}%
_{l+1}^{\prime }\right) }\frac{\left( \mathbf{P}_{l}^{\prime }\mathbf{P}%
_{l+1}^{\prime }.\mathbf{PQ}_{0}\right) }{\left( \mathbf{P}_{l}^{\prime }%
\mathbf{P}_{l+1}^{\prime }.\mathbf{P}_{l}^{\prime }\mathbf{P}_{l+1}^{\prime
}\right) \left\vert \mathbf{PQ}_{0}\right\vert }  \nonumber \\
&&\times \left( \left( \mathbf{P}_{l}^{\prime }\mathbf{P}_{l+1}^{\prime }.%
\mathbf{PS}_{1}\right) -\left( \mathbf{P}_{l}^{\prime }\mathbf{P}%
_{l+1}^{\prime }.\mathbf{PS}_{2}\right) \right) ^{2}  \label{a1.8}
\end{eqnarray}%
where $S_{1},S_{2}$ are arbitrary points of the space-time. Summation is
produced over all world lines of particles perturbing the space-time
geometry. The segment $\mathbf{P}_{l}^{\prime }\mathbf{P}_{l+1}^{\prime }$
is infinitesimal element of the world line $\mathcal{L}_{\left( s\right) }$
of one of perturbing particles. The point $P_{l}^{\prime }$ is near to the
point $P$, which is a middle of the segment $S_{1}S_{2}$.%
\begin{equation}
\sigma (P,P_{l}^{\prime })=0,\qquad \mathbf{PS}_{1}=-\mathbf{PS}_{2}
\label{a1.9}
\end{equation}%
The vectors $\mathbf{PQ}_{i},$ $i=0,1,2,3$ are basic vectors at the point $P$%
. Vector $\mathbf{PQ}_{0}$ is timelike. If $\sigma _{0}$ is the unperturbed
world function of space-time geometry without particles, then $\sigma
=\sigma _{0}+\delta \sigma $ is the world function of the space-time
geometry after appearance of perturbing particles. One should use the world
function $\sigma $ at calculation of scalar products in rhs of (\ref{a1.8})
by the formula (\ref{a1.3}). At first the world function $\sigma $ is
unknown, and relation (\ref{a1.8}) is an equation for determination of $%
\sigma $.

Equation (\ref{a1.8}) is solved by the method of subsequent approximations.
At the first step one calculates rhs of (\ref{a1.8}) by means of $\sigma
_{0} $ and obtains $\sigma _{1}=\sigma _{0}+\delta \sigma _{0}.$ At the
second step one calculates rhs of (\ref{a1.8}) by means of $\sigma _{1}$ and
obtains $\sigma _{2}=\sigma _{0}+\delta \sigma _{1}$ and so on.

Applying relation (\ref{a1.8}) to heavy pointlike particle, one obtains in
the first approximation%
\begin{equation}
\sigma _{1}\left( t_{1},\mathbf{y}_{1};t_{2},\mathbf{y}_{2}\right) =\frac{1}{%
2}\left( 1-\frac{4Gm}{c^{2}\left\vert \mathbf{y}_{2}+\mathbf{y}%
_{1}\right\vert }\right) c^{2}\left( t_{2}-t_{1}\right) ^{2}-\frac{1}{2}%
\left( \mathbf{y}_{2}-\mathbf{y}_{1}\right) ^{2}  \label{a1.10}
\end{equation}%
where $m$ is the mass of the particle.

Space-time geometry appears to be non-Riemannian already at the first
approximation, although the metric tensor has the form, which it has in the
conventional gravitation theory for a slight gravitational field. The next
approximations do not change the situatition.

Thus, the space-time geometry appears to be non-Riemannian. Furthermore,
supposition on the Riemannian space-time leads to an ambiguity of the world
function for large difference of times $(t_{1}-t_{2})$ even in the case of a
gravitational field of a heavy particle. It is conditioned by the fact, that
there are many geodesics, connecting two points. It is forbidden in a
physical geometry, where the world function must be single-valued.

Thus, generalization of the relativity theory on the general case of the
space-time geometry is generated by our progress in geometry and by a use of
adequate relativistic concepts. The deformation principle is not a
hypotheses, but it is the principle, which lies in the basis of physical
geometry. The uniform formalism, suitable for both continuous and discrete
geometries, is characteristic for physical geometries. This formalism uses
dynamic equations in the form of finite difference equations. Sometimes
these equations have a form of finite relations. The uniform formalism is
formulated in coordinateless form. It gets rid of necessity to consider
coordinate transformations and their invariants.

\section{Particle dynamics in physical space-time \newline
geometry}

The contemporary elementary particle theory (EPT) is qualified usually as
the elementary particle physics (EPP). However, it should be qualified more
correctly as an elementary particle chemistry (EPC). The fact is that, the
structure of the elementary particle theory reminds the periodical system of
chemical elements. Both conceptions classify elementary particles (and
chemical elements). On the basis of the classification both conceptions
predict successfully new particles (and chemical elements). Both conceptions
are axiomatic (but not model) constructions.

The periodical system of chemical elements has given nothing new for
investigation of the atomic structure of chemical elements. One should not
expect any information about elementary particle structure from contemporary
EPT. For this purpose a model approach to EPT is necessary.

The simplest particle is considered usually as a point in usual 3D-space.
This point is equipped by a mass and by a momentum 4-vector. One may to
prescribe an electric charge and some other characteristics to the point.
The aggregate of this information forms a nonrelativistic concept of a
particle. This concept of a particle is based on the concept of the linear
vector space, \textit{which is based in turn on the concept of axiomatizable
continuous space-time geometry}.

In the consecutive relativistic theory one should use another concept of a
particle. The simplest particle is defined by two points $P,$ $P^{\prime }$
in the space-time. The vector $\mathbf{PP}^{\prime }$, formed by the two
points, is a geometric momentum of the particle. Its length $\mu =\left\vert 
\mathbf{PP}^{\prime }\right\vert $ is the geometric mass of the particle.
The geometric mass $\mu $ and momentum $\mathbf{PP}^{\prime }\mathbf{\ }$are
connected with conventional mass $m$ and 4-momentum $\mathbf{p}$ by means of
relations%
\begin{equation}
m=b\mu ,\qquad \mathbf{p}=bc\mathbf{PP}^{\prime }  \label{a2.1}
\end{equation}%
where $b$ is some universal constant, and $c$ is the speed of the light. The
electric charge appears in the 5D-geometry of Kaluza-Klein as a projection
of 5-momentum on the additional fifth dimension, which is a chosen
direction.\ Projection on this direction is invariant, because the direction
is chosen. As a result all parameters of a particle appear to be
geometrized. A free motion of the simplest particle in the properly chosen
5D-geometry of the space-time is equivalent to motion of a charged particle
in the given gravitational and electromagnetic fields of the Minkowskian
space-time geometry. \textit{Such a concept of a particle may be used in any
space-time geometry (nonaxiomatizable and discrete)}.

A particle may have a complicated structure, In this case the particle is
described by its skeleton $\mathcal{P}_{n}=\left\{
P_{0},P_{1},...P_{n}\right\} $, consisting of $n+1$ space-time points $%
n=1,2,...$The question: "What does unite the skeleton points in a particle"
is relevant only in the space-time geometry with unlimited divisibility. In
the physical geometry \cite{R2001,R2007,R2008a} the skeleton points may be
connected between themselves simply as points of a geometry with a limited
divisibility.

The particle evolution is described by a chain $\mathcal{C}$ of connected
skeletons \cite{R2008a,R2010a}.%
\begin{equation}
\mathcal{C}=\dbigcup\limits_{s}\mathcal{P}_{n}^{(s)}  \label{a2.1a}
\end{equation}%
Adjacent skeletons of the chain are equivalent.%
\begin{equation}
\mathcal{P}_{n}^{\left( s+1\right) }\text{eqv}\mathcal{P}_{n}^{\left(
s\right) }:\quad \mathbf{P}_{i}^{(s+1)}\mathbf{P}_{k}^{(s+1)}\text{eqv}%
\mathbf{P}_{i}^{(s)}\mathbf{P}_{k}^{(s)}\quad i,k=0,1,...n,\quad s=...0,1,...
\label{a2.2}
\end{equation}%
Points $P_{1}^{\left( s\right) }$and $P_{0}^{\left( s+1\right) }$of the
chain coincide $s=...0,1,..$ Then according to (\ref{a2.2}) the leading
vector $\mathbf{P}_{0}^{\left( s\right) }\mathbf{P}_{1}^{\left( s\right) }=%
\mathbf{P}_{0}^{\left( s\right) }\mathbf{P}_{0}^{\left( s+1\right) }$ of
skeleton $\mathcal{P}_{n}^{(s)}$ is equivalent to the leading vector $%
\mathbf{P}_{0}^{\left( s+1\right) }\mathbf{P}_{1}^{\left( s+1\right) }=%
\mathbf{P}_{0}^{\left( s+1\right) }\mathbf{P}_{0}^{\left( s+2\right) }$ of
skeleton $\mathcal{P}_{n}^{(s+1)}$, i.e.%
\begin{equation}
\mathbf{P}_{0}^{\left( s\right) }\mathbf{P}_{0}^{\left( s+1\right) }\text{eqv%
}\mathbf{P}_{0}^{\left( s+1\right) }\mathbf{P}_{0}^{\left( s+2\right) }
\label{a2.3}
\end{equation}

In the explicit form equations (\ref{a2.1a}), (\ref{a2.2}), describing the
world chain, look as follows%
\begin{eqnarray}
(\mathbf{P}_{i}^{(s+1)}\mathbf{P}_{k}^{(s+1)}.\mathbf{P}_{i}^{(s)}\mathbf{P}%
_{k}^{(s)}) &=&\left\vert \mathbf{P}_{i}^{(s+1)}\mathbf{P}%
_{k}^{(s+1)}\right\vert \cdot \left\vert \mathbf{P}_{i}^{(s)}\mathbf{P}%
_{k}^{(s)}\right\vert ,  \label{a2.4} \\
\left\vert \mathbf{P}_{i}^{(s+1)}\mathbf{P}_{k}^{(s+1)}\right\vert
&=&\left\vert \mathbf{P}_{i}^{(s)}\mathbf{P}_{k}^{(s)}\right\vert ,\quad
P_{1}^{\left( s\right) }=P_{0}^{\left( s+1\right) }  \label{a2.5} \\
i,k &=&0,1,...n,\quad s=...0,1,...  \nonumber
\end{eqnarray}%
where scalar products are defined via world functions by the relation (\ref%
{a1.3}).

Rotation of a skeleton is absent. The translational motion is carried out
along the leading vector $\mathbf{P}_{0}\mathbf{P}_{1}$. Dynamics is
described by means of finite difference equations. It is reasonable, if the
space-time geometry may be discrete. The leading vector describes the
evolution direction in the space-time.

The number of dynamic equations is equal to $n(n+1)$, whereas the number of
variables to be determined is equal to $Nn$. Here $N$ is the dimension of
the space-time, and $n+1$ is the number of points in the particle skeleton.
The difference between the number of equations and the number of variables,
which are to be determined, may lead to different results.

\begin{enumerate}
\item Multivariance, i.e. ambiguity of the world chain links position, when $%
n(n+1)<Nn$. It is characteristic for simple skeletons, which contain small
number of points. Multivariance is responsible for quantum effects \cite{R91}%
.

\item Zero-variance, i.e. absence of solution of equations, when $n(n+1)>Nn$%
. It is characteristic for complicated skeletons, which contain many points.
Zero-variance means a discrimination of particles with complicated
skeletons. As a result there exist only particles, having only certain
values of masses and other parameters.
\end{enumerate}

\textit{Quantum indeterminacy and discrimination mechanism are two different
sides of the particle dynamics}. The conventional theory of elementary
particles has not a discrimination mechanism, which could explain a discrete
spectrum of masses.

There are two sorts of elementary particles: bosons and fermions. Boson has
not its own angular momentum (spin). It is rather reasonable, because motion
of elementary particles is translational. However, the fermions have a
discrete spin, which looks rather unexpected at the translation motion. Spin
of a fermion appears as a result of translation motion along a space-like
helix with timelike axis \cite{R2004a,R2004b,R2008}. The helix world line of
a free particle is possible only for spacelike world line. It is conditioned
by multivariance \cite{R2008c} of the space-time geometry with respect to
spacelike vectors. This multivariance takes place even for space-time of
Minkowski. This multivariance takes place for any space-time geometry. It
does not vanish in the limit $\hbar \rightarrow 0$.

However, in the space-time geometry of Minkowski the helix world chain is
impossible, because the temporal component of momentum increases infinitely.
For existence of the helix world chain, the world function $\sigma $ is to
have the form%
\begin{equation}
\sigma =\left\{ 
\begin{array}{ccc}
f\left( \sigma _{\mathrm{M}}\right) & \text{if} & \left\vert \sigma _{%
\mathrm{M}}\right\vert <\sigma _{0} \\ 
\sigma _{\mathrm{M}}+\lambda _{0}^{2}\text{sgn}\left( \sigma _{\mathrm{M}%
}\right) & \text{if} & \left\vert \sigma _{\mathrm{M}}\right\vert >\sigma
_{0}%
\end{array}%
\right. ,\qquad \lambda _{0}^{2}=\frac{\hbar }{2bc},\qquad \sigma _{0}=\text{%
const}  \label{a2.7}
\end{equation}%
\begin{equation}
\left\vert f\left( \sigma _{\mathrm{M}}\right) \right\vert <\left\vert
\sigma _{\mathrm{M}}\right\vert \frac{\sigma _{0}+\lambda _{0}^{2}}{\sigma
_{0}},\qquad \left\vert \sigma _{\mathrm{M}}\right\vert <\sigma _{0}
\label{a2.8}
\end{equation}

In the conventional relativity theory the helix spacelike world lines are
not considered, because one assumes, that they are forbidden by the
relativity principles. Fermions are described usually by means of the Dirac
equation, which needs introduction of such special quantities as $\gamma $%
-matrices. A use of $\gamma $-matrices generates a mismatch between the
particle velocity and its mean momentum. (The quantum mechanics uses the
mean momentum always \cite{R2004}.) This enigmatic mismatch is explained
easily by means of the helix world chain. The velocity is tangent to helix,
whereas the mean momentum is directed along the axis of helix.

Besides, the fermion skeleton is to contain not less, than three points. It
is necessary for stabilization of the helix world line \cite{R2004a,R2004b}.
Existence of the fermion is possible only at certain values of its mass,
which depends on the space-time geometry (the form of function $f$ in (\ref%
{a2.7})) and on a choice of the skeleton points.

Thus, the spin and magnetic moment of fermions appear to be connected with
spacelike world chain and with multivariance of the space-time geometry with
respect to space-like vectors. At the conventional approach to geometry the
spacelike world lines are considered to be incompatible with the relativity
principles. Spin is associated with existence of enigmatic $\gamma $%
-matrices. Multivariance with respect to timelike vectors is slight (it
vanishes in the limit $\hbar =0$). Multivariance with respect to spacelike
vectors is strong (it is not connected with quantum effects)

The particle motion is free in the properly chosen space-time geometry.
However, the particle motion can be described in arbitrary geometry, given
on the same point set, where the true geometry is given. The world function $%
\sigma $ of the true geometry is presented in the form%
\begin{equation}
\sigma \left( P,Q\right) =\sigma _{K_{0}}\left( P,Q\right) +d\left(
P,Q\right)  \label{a2.9}
\end{equation}%
where $d\left( P,Q\right) $ is some addition to the world function of $%
\sigma _{K_{0}}\left( P,Q\right) $ of the space-time geometry of
Kaluza-Klein, which is used in the given case as a basic geometry. In this
geometry the particle motion ceases to be free. It turns into a motion in
force fields, whose form is determined by the form of addition $d\left(
P,Q\right) $.

Progress in the elementary particle dynamics is conditioned by a progress in
geometry and by a use of adequate relativistic concepts. The suggested
elementary particle dynamics is a model conception. It is demonstrable and
simple. Multivariance of the geometry explains freely quantum effects. The
zero-variance generates a discriminational mechanism, responsible for
discrete characteristics of elementary particles. Mathematical technique is
formulated in a coordinateless form, that gets rid of a necessity to
investigate coordinate transformations and their invariants. Two-point
technique of the dynamics and many-point skeletons contain a lot of
information, which should be only correctly ordered. Simple principles of
dynamics reduce a construction of the elementary particle theory to formal
calculations of different skeletons dynamics at different space-time
geometries. There is a hope, that true skeletons of elementary particles can
be obtained by means of the discrimination mechanism of the true space-time
geometry. At any rate, having been constructed in the framework of simple
dynamic principles, this dynamics explains freely discrete spins and
discrete masses of fermions and mismatch between the particle velocity and
its mean momentum. These properties are described usually by introduction of 
$\gamma $-matrices, that is a kind of fitting.


\begin{thebibliography}{99}
\bibitem{R2010} Yu. A. Rylov, Logical reloading as overcoming of crisis in
geometry. \textit{e-print 1005.2074 }

\bibitem{R2001} Yu.A.Rylov, Geometry without topology as a new conception of
geometry. \textit{Int. Jour. Mat. \& Mat. Sci.} \textbf{30}, iss. 12,
733-760, (2002).

\bibitem{R2007} Yu.A.Rylov, Non-Euclidean method of the generalized geometry
construction and its application to space-time geometry in \textit{Pure and
Applied Differential geometry} pp.238-246. eds. Franki Dillen and Ignace Van
de Woestyne. Shaker Verlag, Aachen, 2007. See also \textit{e-print
Math.GM/0702552.}

\bibitem{S60} J.L.Synge, \textit{Relativity: the General Theory. }Amsterdam,
North-Holland Publishing Company, 1960.

\bibitem{R91} Yu.A.Rylov, Non-Riemannian model of the space-time responsible
for quantum effects. \textit{Journ. Math. Phys}. \textbf{32(8)}, 2092-2098,
(1991).

\bibitem{R2009} Yu. A. Rylov, Relativistic nearness of events and
deformation principle as tool of the relativity theory generalization on the
arbitrary space-time geometry. \textit{e-print 0910.3582v4}

\bibitem{R2008a} Yu. A. Rylov, Generalization of relativistic particle
dynamics on the case of non-Riemannian space-time geometry. \textit{Concepts
of Physics} \textbf{6}, iss.4, 605, (2009). Se also \textit{e-print 0811.4562%
}.

\bibitem{R2010a} Yu. A. Rylov, Necessity of the general relativity revision
and free motion of particles in non-Riemannian space-time geometry. \textit{%
e-print 1001.5362v1.}

\bibitem{R2004a} Yu. A. Rylov, Is the Dirac particle composite? \textit{%
eprint physics/0410045}

\bibitem{R2004b} Yu. A. Rylov, Is the Dirac particle completely
relativistic? \textit{e-print} \textit{physics/0412032}.

\bibitem{R2008} Yu. A. Rylov, Geometrical dynamics: spin as a result of
rotation with superluminal speed. \textit{e-print 0801.1913.}

\bibitem{R2008c} Yu. A. Rylov, Multivariance as a crucial property of
microcosm. \textit{Concepts of Physics} \textbf{5}, iss.1, 89 -117, (2009).
See also \textit{e-print 0806.1716.}

\bibitem{R2004} Yu. A. Rylov, Hydrodynamical interpretation of quantum
mechanics: the momentum distribution \textit{e-print} \textit{physics/0402068%
}.
\end{thebibliography}
\end{document}